# Estudo e Implementação de Algoritmos de Roteamento sobre Grafos em um Sistema de Informações Geográficas

Rudini Menezes Sampaio[1]
Horácio Hideki Yanasse[2]

[1]UFLA – Universidade Federal de Lavras
DCC – Departamento de Ciência da Computação
rudini@comp.ufla.br

[2]INPE – Instituto Nacional de Pesquisas Espaciais
LAC - Laboratório Associado de Computação e Matemática Aplicada
horacio@lac.inpe.br

**RESUMO**
Este artigo discute alguns dos principais algoritmos de roteamento em grafos, como menor caminho, árvore de custo mínimo, carteiro chinês e caixeiro viajante, e apresenta sua implementação em um Sistema de Informações Geográficas.
**Palavras-Chave.** Pesquisa Operacional, Algoritmos de Roteamento, Teoria dos Grafos.

## 1. Introdução

Nesse documento, apresenta-se um estudo de alguns dos principais problemas de roteamento em grafos, como Menor Caminho, Árvore de Custo Mínimo, Carteiro Chinês e Caixeiro Viajante, bem como o desenvolvimento e a implementação em um Software gráfico e prático de algoritmos que os solucionem em tempo hábil.

Descreve-se ainda o desenvolvimento de um aplicativo que permite a visualização e manipulação de grafos sobre arquivos de figura e a execução dos algoritmos desenvolvidos sobre diversos contextos, possibilitando o seu uso em casos de Fotos de Satélites, GPS, Mapas e outros.

## 2. Menor caminho

O problema do menor caminho é bastante conhecido e tem como objetivo obter um percurso mínimo entre dois ou mais vértices de um grafo. Neste caso, um grafo pode representar uma malha rodoviária, distâncias geográficas e etc...

Além de ser clara a aplicabilidade dessas técnicas para o problema de motoristas, correios ou serviços de emergência, existe uma motivação muito mais importante para se começar a estudá-los: a determinação dos menores caminhos aparece constante e consistentemente como um subproblema de problemas mais complexos em grafos.

### 2.1 Algoritmo de Dijkstra: Entre um dado ponto e os demais

No algoritmo de Dijkstra, o objetivo é obter o menor caminho entre um dado vértice fixo e todos os demais vértices do grafo. Por exemplo, saber a distância mínima de São Paulo para todas as cidades do Estado.

O algoritmo consiste basicamente em fazer uma visita por todos os nós do grafo, iniciando no nó fixo dado e encontrando sucessivamente o nó mais próximo, o segundo mais próximo, o terceiro mais próximo e assim por diante, um por vez, até que todos os nós do grafo tenham sido visitados.

Na figura 1, exemplificamos a execução do algoritmo de Dijkstra no Sistema, com um grafo sobre um mapa de uma cidade, cujas arestas representam as ruas.

### 2.2 Algoritmo de Floyd: Entre todos os pares de pontos

É muito comum o caso de ser necessário o menor caminho entre todos os pares de nós de um grafo. Como exemplo podemos citar a preparação de tabelas indicando distâncias entre todas as cidades em mapas rodoviários de estados ou regiões, ou obter o menor caminho que parta de um nó dado, passe por alguns nós intermediários dados em ordem de prioridade e chegue em um nó final. Possui muitas aplicações, como entrega de encomendas em hora marcada por empresas distribuidoras de produtos, etc.

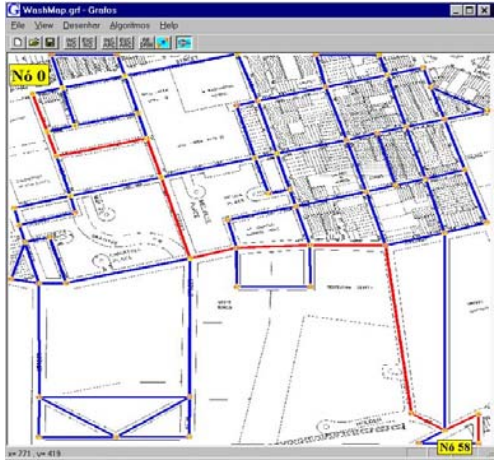
Figura 1: Algoritmo de Dijstra sobre Mapas

Uma solução óbvia é repetir o algoritmo de Dijkstra sucessivamente para todos os nós do grafo. Uma solução mais eficiente é conhecida como *algoritmo de Floyd*, que utiliza programação dinâmica.

A idéia geral desse algoritmo é atualizar a matriz de menores distâncias **n** vezes (onde **n** é o número de nós do grafo) procurando na **K**-ésima interação por melhores distâncias entre pares de nós que passem pelo vértice **K**.

Nas figuras 2, 3 e 4, exemplificamos a execução do algoritmo de Floyd no Sistema, com um grafo com as possíveis jogadas do Cavalo no tabuleiro de xadrez, um grafo com uma fotografia de satélite, onde as arestas representam as rodovias, e outro grafo com o mapa da cidade de São José dos Campos, em São Paulo, onde as arestas representam as ruas.

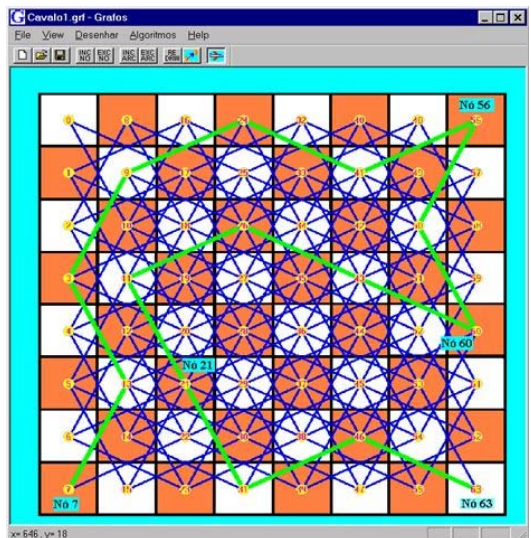
Figura 2: Algoritmo de Floyd sobre o Tabuleiro

Execução em **verde** para a seqüência de nós 7 56 60 21 63, sobre o tabuleiro.

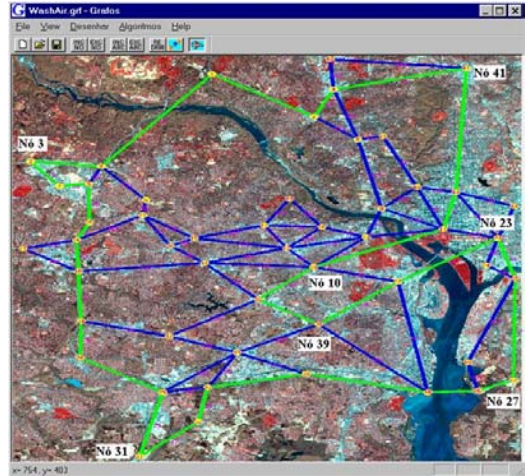
Figura 3: Algoritmo de Floyd sobre Fotos de Satélite

Execução em **verde** para a seqüência de nós 3, 31, 27, 23, 39, 10, 41 e 3, sobre a foto de satélite

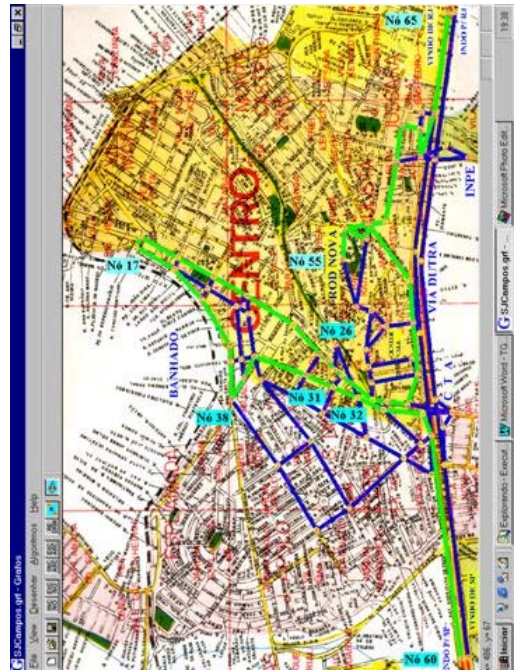
Figura 4: Algoritmo de Floyd sobre Mapas

Execução em **verde** para a seqüência de nós 65-55-17-38-60, representando os pontos turísticos da cidade.

## 3. Árvore de custo mínimo (Algoritmo de Prim)

O Problema da Árvore de Custo Mínimo é encontrar uma árvore que contenha todos os vértices do grafo e cuja soma das arestas seja mínima.

Este problema aparece, por exemplo, no seguinte contexto: é dado um mapa de $n$ cidades rurais com uma matriz listando as distâncias euclidianas entre todos os pares possíveis de cidades e deseja-se obter o menor comprimento de rodovias necessárias para unir todas elas. Outro contexto importante seria para auxiliar na decisão de onde se localizar postos de emergência ou delegacias de polícia em uma cidade, por exemplo. Além dessas aplicações, a solução desse problema é de grande utilidade para a solução de outros problemas mais complexos em grafos, tais como o do caixeiro viajante.

O Algoritmo de Prim é um algoritmo "*guloso*" cuja idéia básica é: Escolhendo um nó arbitrário inicialmente, visite todos os nós do grafo escolhendo como próximo a ser visitado o nó mais "***perto***" de um dos vértices já visitados.

Nas figuras 5 e 6, exemplificamos a execução do algoritmo de Prim no Sistema, com um grafo sobre um mapa de uma cidade, onde as arestas representam as ruas, e outro grafo com uma fotografia de satélite, onde as arestas representam as rodovias.

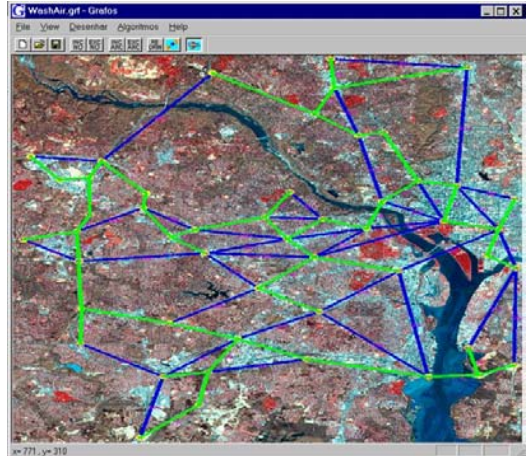
Figura 6: Algoritmo de Prim sobre Fotos de Satélite

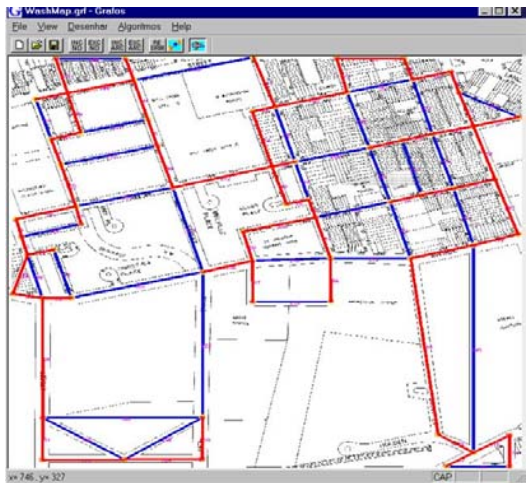
Figura 5: Algoritmo de Prim sobre Mapas

## 4. Carteiro Chinês

Considere o caso de um carteiro responsável pela correspondência de uma área da cidade. O carteiro deverá sempre começar o percurso em um nó inicial (os Correios), deverá passar por todas as ruas (arestas) e retornar ao nó inicial.

A questão mais natural a se perguntar é: a fim de minimizar a distância total que o carteiro percorre, como deverá ser a sua rota de forma que ele passe por todas as ruas ao menos uma vez ?

Essa questão é conhecida como o Chinese Postman Problem, nome derivado do fato de ter sido no jornal Chinese Mathematics em 1952 a primeira vez em que esse problema foi discutido.

A história desse problema é bastante interessante. No século XVIII, os moradores da cidade russa de Königsberg queriam realizar um desfile que pudesse passar pelas sete (Figura 7) pontes sobre o rio Prevel apenas uma vez e passaram o problema para o matemático suíço *Leonhard Euler* resolver.

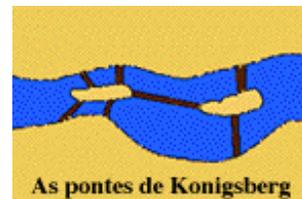
Figura 7: Pontes de Königsberg

*Euler* provou que não existe solução para esse problema e que, para isso, nenhum vértice do grafo pode ter grau ímpar. Assim, para se realizar um ciclo euleriano em um grafo, este deverá ser modificado de modo a tornar de grau par todos os seus nós de grau ímpar.

Para isso, é necessário combinar dois a dois todos os seus nós de grau ímpar. Esse problema de combinação é chamado de "*Pairwise Matching*", e foi resolvido por Edmonds. Em alto nível, o algoritmo de solução do carteiro chinês é:

**PASSO 1**: Seja **M** o conjunto de todos os nós de grau ímpar do grafo. Digamos que existam **m** deles.

**PASSO 2:** Encontre a combinação de pares de nós de **M** cuja soma das distâncias seja mínima.

**PASSO 3:** Encontre as arestas dos menores caminhos entre os dois nós que compõem cada um dos **m**/2 pares obtidos no passo 2.

**PASSO 4:** Para cada um dos pares obtidos em passo 2, adicione ao grafo as arestas obtidos no passo 3. O grafo resultante não contém nó de grau ímpar.

**PASSO 5:** Encontre um Ciclo Euleriano, que é a solução para o problema do carteiro chinês.

### 4.1 Ciclo Euleriano

O ciclo Euleriano é um percurso que começa em um nó de partida, passa por todos os nós do grafo e termina no nó de partida. A obtenção do Ciclo Euleriano depende da execução do algoritmo de "*Matching*", pois o grafo não pode conter nós de grau ímpar.

**PASSO 1:** Se não existem nós adjacentes ao nó de partida, então PARE. O Ciclo euleriano já foi descrito. Caso contrário, vá para PASSO 2.

**PASSO 2:** Dos nós adjacentes ao nó de partida escolha um cuja aresta que o ligue ao nó de partida não seja uma Ponte (ou seja, a sua eliminação não torna o grafo desconexo) e vá para ele. Elimine do grafo a aresta entre esses nós. Faça o nó de partida ser esse novo nó escolhido. Vá para PASSO 2.

O problema em si está no conceito de aresta Ponte ou Bridge. Para sabermos se uma aresta é ponte ou não, deveremos eliminá-la do grafo e verificar se o grafo resultante é ou não conexo.

Na figura 8, exemplificamos a execução do algoritmo do Carteiro Chinês no Sistema, com um grafo sobre uma fotografia de satélite, onde as arestas representam rodovias. Note que algumas arestas estão duplicadas, pelo *matching*.

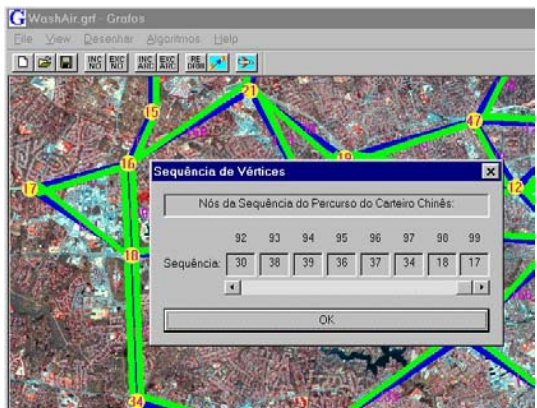
Figura 8: Carteiro Chinês sobre Fotos de satélite

### 5. Caixeiro Viajante

De todos os problemas de cobrimento de nós, o mais conhecido é o Problema do Caixeiro Viajante, cujo objetivo é encontrar a rota de menor distância que inicie em um dado nó de um grafo, visite todos os membros de um conjunto específico de nós do grafo uma única vez e retorne ao nó inicial.

Para esse problema, assumimos uma particularização muito utilizada, que é quando o grafo é totalmente conectado e satisfaz a desigualdade triangular. Mesmo assim, esse problema ainda é NP-Difícil. Assim, ao invés de buscar algoritmos exatos, é mais interessante tentar obter algoritmos polinomiais aproximativos.

O algoritmo aproximativo utilizado foi desenvolvido por *Christofides* em 1976, que provou que o percurso obtido não é maior que 50% do ótimo. O algoritmo, descrito a seguir, pode ainda ser melhorado adicionando as otimizações OPT2 e OPT3.

**PASSO 1:** Encontre a Árvore **T** de custo mínimo do grafo.

**PASSO 2:** Seja **n₀** o número de nós de grau ímpar dos **n** nós de **T** (**n₀** é par). Encontre o *matching* mínimo dos pares desses **n₀** nós (*Edmonds*). Seja **M** o grafo das arestas da solução do *matching*. Seja **H=M∪T**, um grafo.

**PASSO 3:** O grafo **H** não contém nós de grau ímpar. Encontre um Ciclo Euleriano sobre **H**, que é a solução aproximada do problema do Caixeiro Viajante.

Na figura 9, exemplificamos a seqüência de execução do algoritmo. Nesse exemplo, o percurso obtido não chega a ser 1% pior que o percurso ótimo.

Nas figuras 10 e 11, exemplificamos a execução do algoritmo aproximativo do Caixeiro Viajante no Sistema, com um grafo semelhante ao da figura 9, e com um grafo sobre o tabuleiro de xadrez.

### 6. Conclusões

Este trabalho foi desenvolvido em um projeto do LAC/INPE voltado para a implementação de algoritmos em Sistemas Geográficos de Informação, como o software ARC-INFO (ESRI), em uso no projeto. Este trabalho foi desenvolvido ainda permitindo a possível introdução de novos algoritmos ao sistema, tornando-se uma boa ferramenta de comparação dos algoritmos de roteamento em grafos.

Outros algoritmos ainda devem ser implementados, como Carteiro Chinês em Grafos Orientados, Rural Postman Problem, Carteiro Chinês com Capacidade da Mochila e Problemas de Localização de Facilidades (**p-medianas**)

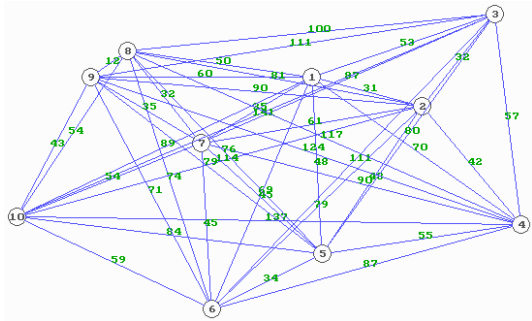

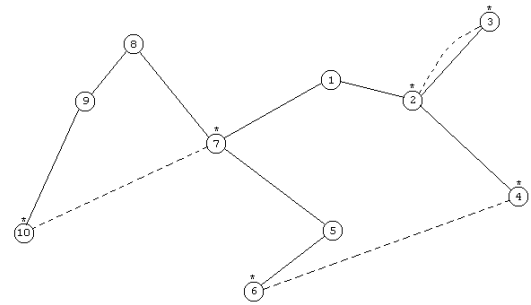

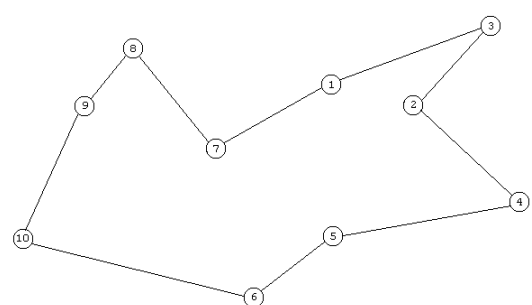

Figura 9: Seqüência de Passos do Caixeiro Viajante

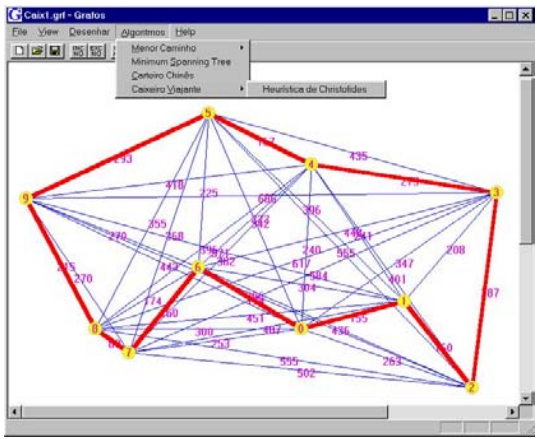

Figura 10: Caixeiro Viajante no Sistema

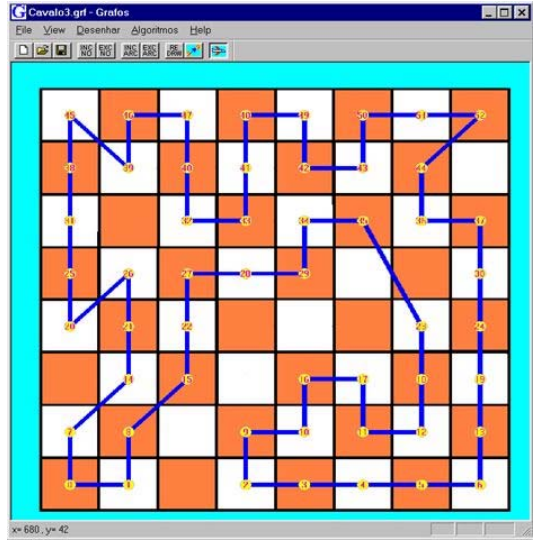

Figura 11: Caixeiro Viajante sobre o Tabuleiro

O Sistema implementado permite ainda a manipulação gráfica da estrutura do grafo, possibilitando a inclusão de novos nós e arestas, bem como o seu reposicionamento, como pode ser visto na figura 12.

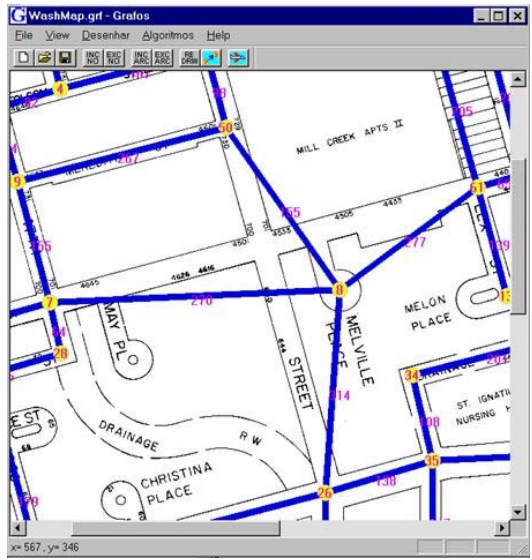

Figura 12: Reposicionamento de vértices no Sistema

O sistema permite ainda a visualização em *zoom* do grafo e da execução dos algoritmos, como pode ser visto na figura 13.

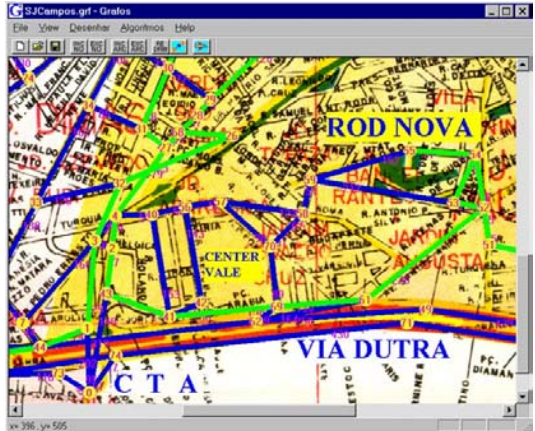

Figura 13: Visualização do grafo em *zoom* no Sistema

## 7. Bibliografia

[1] Cormen, T.; Leiserson, C.; Rivest, R., Introduction to Algorithms, MIT Press, 2° edição, 2001.

[2] Papadimitriou, C.; Steiglitz, K., Combinatorial Optimization – Algorithms and Complexity, Dover Publications, Mineola, New York, 1998.

[3] Aho, A.V.; Hopcroft, J.E.; Ullman, J.D., *The Design and Analysis of Computer Algorithms*, Addison-Wesley, 1974.

[4] Larson, R.C.; Odoni, A.R., *Urban Operations Research*, Prentice Hall, 1981.

[5] Campello, R.E.; Maculan, N., *Algoritmos e heurísticas*, EDUFF, 1994.